\documentclass[12pt]{article}
\usepackage{bm}
\usepackage{booktabs,microtype,siunitx,tabularx}
\usepackage[margin=1.5in]{geometry}

\usepackage{natbib}
\setcitestyle{authoryear,open={(},close={)}}
\usepackage{mathtools}
\usepackage{amsfonts}
\usepackage{listings}
\usepackage{amsmath}
\renewcommand*{\mathbf}[1]{\ifmmode\bm{#1}\else\textbf{#1}\fi}

\title{A Novel Nonlinear Fertility Catastrophe  Model Based on Thom's Differential Equations of Morphogenesis}

\author{Rolando Gonzales Martinez \thanks{University of Groningen, Population Research Centre}}

\begin{document}

\maketitle

\begin{abstract}
A novel fertility model based on Thom's nonlinear differential equations of morphogenesis is presented, utilizing a three-dimensional catastrophe surface to capture the interaction between latent non-catastrophic fertility factors and catastrophic shocks. The model incorporates key socioeconomic and environmental variables and is applicable at macro-, meso-, and micro-demographic levels, addressing global fertility declines, regional population disparities, and micro-level phenomena such as teenage pregnancies. This approach enables a comprehensive analysis of reproductive health at aggregate, sub-national, and age-group-specific levels. An agent-based model for teenage pregnancy is described to illustrate how latent factors---such as education, contraceptive use, and parental guidance---interact with catastrophic shocks like socioeconomic deprivation, violence, and substance abuse. The bifurcation set analysis shows how minor shifts in socioeconomic conditions can lead to significant changes in fertility rates, revealing critical points in fertility transitions. By integrating Thom's morphogenesis equations with traditional fertility theory, this paper proposes a groundbreaking approach to understanding fertility dynamics, offering valuable insights for the development of public health policies that address both stable fertility patterns and abrupt demographic shifts.
\end{abstract}

\section{Introduction}

The global decline in fertility rates is a demographic shift with profound implications for social systems and economies worldwide. This decline, driven by factors such as rising levels of female education, economic development, urbanization, and improved living standards, presents challenges in countries like Japan and South Korea, where fertility rates have fallen well below replacement levels. While the characterization of these declines as catastrophic varies depending on the context, their impact on aging populations and labor force contraction is widely acknowledged.

In many societies, shifts in values and attitudes toward family size, delays in marriage, and increased gender equality have contributed to these changes. Moreover, advances in healthcare and reductions in child mortality have further altered fertility patterns. These factors interact in complex ways, and their relative importance varies across regions, underscoring the need for a nuanced understanding of fertility dynamics to inform policy interventions.

Historically, mathematical models of fertility have focused on estimating fertility rates \citep{brass1975general}, forecasting trends \citep{lutz1999forecasting, lee1992modeling}, analyzing the quantum and tempo of fertility \citep{bongaarts1998estimating}, and modeling age-specific fertility patterns \citep{asili2014using, gaire2022mathematical}. This study proposes a novel fertility model based on Thom's nonlinear differential equations of morphogenesis. The model integrates traditional fertility factors with catastrophic socioeconomic and environmental shocks that influence pregnancy rates and, on a larger scale, fertility outcomes. In the fertility catastrophe model, the system states are influenced by control parameters that can induce abrupt changes in fertility outcomes. These parameters govern the transition from stable fertility trends to sudden, discontinuous shifts in fertility patterns.

Of particular interest in the model are the bifurcation parameters, which determine the discontinuities in the fertility response surface. When these parameters exceed certain thresholds, they may trigger catastrophic events, resulting in marked decreases or increases in pregnancy rates. The estimation and testing of these parameters provide critical insights into the occurrence and impact of catastrophic fertility events. Importantly, the model accounts for both negative and positive fertility shocks, allowing for the examination of population declines as well as potential booms. A thorough understanding of these events can assist policymakers in preparing for demographic shifts, facilitating more robust public health planning and policy responses.

Next section describes the nonlinear fertility catastrophe model and Section \ref{sec:ABM} shows an application of the nonlinear fertility catastrophe model based on agent based mathematical model of teenage pregnancy. Section \ref{sec:discussion} discusses potential applications and Section \ref{sec:conc} concludes. The Python code to replicate the results of the simulations of the cusp catastrophe-agent based model of teenage pregnancy is in the appendix at the end of this document. 

The next section outlines the nonlinear fertility catastrophe model in detail. Section \ref{sec:ABM} shows its application within an agent-based mathematical framework for modeling teenage pregnancy. In Section \ref{sec:discussion}, other potential applications of the model are discussed, and Section \ref{sec:conc} provides concluding observations. The Python code used to replicate the simulation results for the cusp catastrophe-agent based model of teenage pregnancy is included in the appendix at the end of this document.

\section{A Nonlinear Fertility Catastrophe Model}\label{sec:NMFC}

The fertility catastrophe model is based on the fold function proposed by \citet{thom1972stabilite}. In the fold function, the states $s$ of $F_{\Psi} (s)$ are defined by the variables $\xi$ and the parameters $\Psi$, hence $F_{\Psi} (s) := F(\xi,\Psi)$ in the cusp model, and the function $F(\xi,\Psi)$ is equal to:
\begin{equation}
    F(\xi,\Psi) = \psi_1 \xi^4 - \psi_2 \xi^2 + \psi_3 \xi
    \label{eq:cuspF}
\end{equation}

In equation \ref{eq:cuspF}, the variable $\xi$ represents the state of the dynamical system, while the control parameters $\Psi = \{\psi_1, \psi_2, \psi_3\}$ influence the dynamics of the system. Changes in latent control parameters $\Psi$ give rise to various geometric forms in the state variable $\xi$. These control parameters capture the dynamic changes of a system from a stable state to a sudden catastrophic situation, as discussed in \citet{renfrew1978trajectory} and \citet{saunders_1980}. The derivative with respect to the time of the states $\xi$ gives rise to the deterministic dynamical equation:
\begin{equation}
\frac{\partial \xi}{\partial t} = - \frac{\partial V (\xi; \alpha,\beta)}{\partial \xi}
\label{eq:partial}
\end{equation}
In equation \ref{eq:partial}, the changes of $\xi$ across time $t$ are captured by the function $V (\xi; \alpha,\beta)$, hence the function $V (\xi; \alpha,\beta)$ is a time-dependent version of the fold function $F(\xi,\Psi)$ in equation \ref{eq:cuspF}, with the control parameters in the original model of Equation \ref{eq:cuspF} now equal to $\psi_1 = 1/4$, $\psi_2 = 1/2\beta$, $\psi_3 = -\alpha$:
\begin{equation}
-V(\xi; \alpha,\beta) = \alpha \xi + \frac{1}{2}\beta \xi^2 - \frac{1}{4}\xi^4,
\label{eq:cusp}
\end{equation}
In the function $V (\xi; \alpha,\beta)$ the control parameters $\alpha$ and $\beta$ affect the dynamics of $\xi$. The equilibrium points are a function of the control parameters $\alpha$ and $\beta$ and are the result of solutions to equation $\alpha + \beta \xi - \xi^3 = 0$, which is equal to the Cardan discriminant $\delta = 27\alpha - 4\beta^3$ \citep{grasman2010fitting}. Cardan's discriminant differentiates between unimodal ($\delta \leq 0$) and bimodal cases ($\delta > 0$).

Equation \ref{eq:cuspF} is a dynamic deterministic equation that serves as the foundation of the catastrophe model. However, to account for the inherent unpredictability of real-world fertility scenarios, a stochastic component is incorporated into this deterministic equation. This enhancement enables the estimation of a catastrophe function using real population data. By combining deterministic and stochastic elements, the dynamical system becomes more comprehensive and better suited for capturing the complexities of real-world fertility scenarios.

The deterministic model \ref{eq:cusp} can be extended to represent a stochastic process if a Wiener white noise term $dW(t)$, associated with a Gaussian probability distribution with a variance $\sigma^2$, is added to $-\partial V(\xi;\alpha,\beta)/\partial \xi$ to obtain the stochastic differential equation:
\begin{equation}
d\xi = -\frac{\partial V(\xi;\alpha,\beta)}{\partial \xi} dt + dW(t).
\end{equation}

\begin{figure}[ht]
\centering
 \includegraphics[width=14cm]{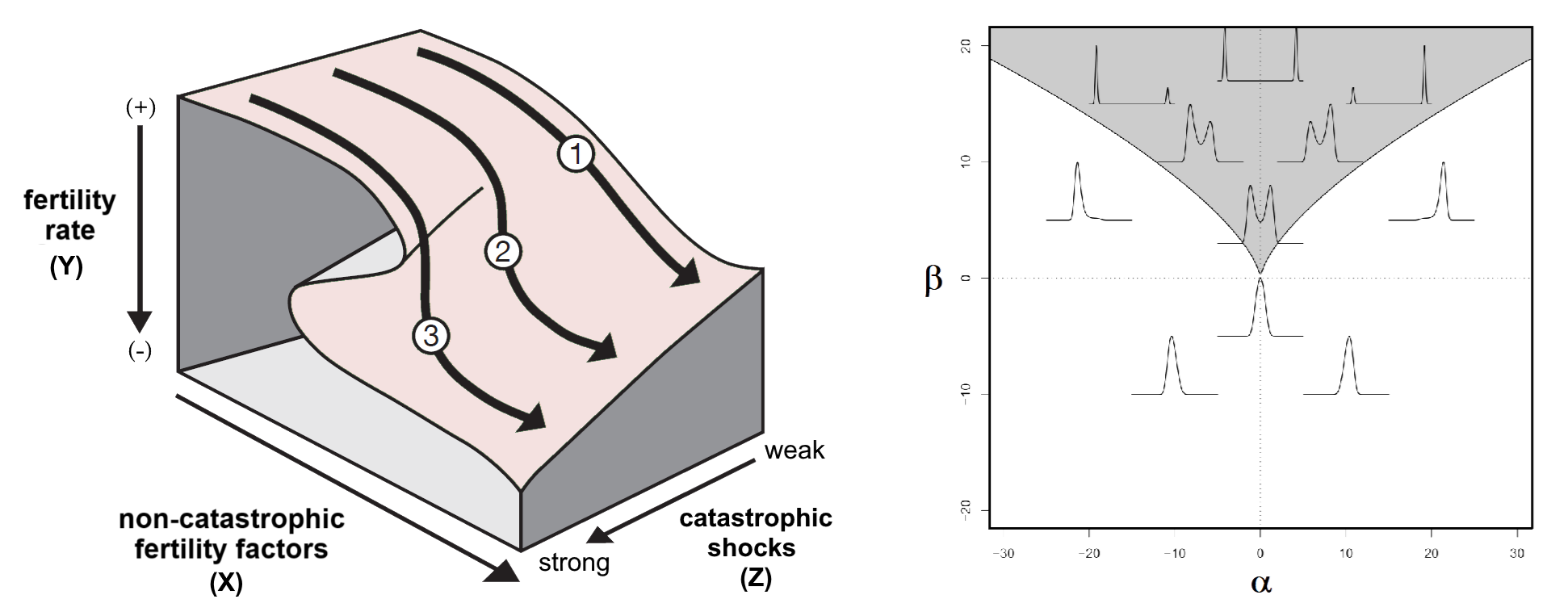} 
 \caption{Left: Response surface of the fertility model based on catastrophic morphogenesis. The model takes into account the effects of $X$ traditional fertility factors, as well as catastrophic shocks ($Z$). Right: state-space of control parameters of the model: $\alpha$ and $\beta$.}
\label{fig1}
\end{figure}

Figure \ref{fig1} (right) shows combinations of values of the control parameters $\alpha$ and $\beta$. The curves in the plane represent the distribution of the number of pregnancies $\xi$ for different values of the control parameters. Unimodality and bimodality of the number of pregnancies $\xi$ is observed for different values of $\alpha$ and $\beta$. The parameter $\alpha \in \mathbb{R}$ controls the asymmetry of the number of pregnancies, in the sense that positive values of $\alpha$ indicate an asymmetry of the distribution towards a non-crisis reproductive environment ($\alpha \to \infty^+$). In contrast, negative values of $\alpha$ indicate an asymmetry of the distribution towards a reproductive crisis ($\alpha \to \infty^-$). The parameter $\beta \in \mathbb{R}$ is a bifurcation parameter that switches the number of pregnancies between a reproductive crisis ($\beta \to \infty^-$) and a catastrophic state of fertility ($\beta \to \infty^+$) caused by a catastrophic event. A non-catastrophic state ($\beta < 0$) is a state of low fertility. In contrast, when $\beta \to \infty^+$, the model captures the impact of catastrophic events, which cause severe, unstable, and catastrophic fertility conditions (Figure \ref{fig1}, right). 

Of particular interest are the values of $\beta \in \mathbb{R}^{0,\infty+}$, which control the discontinuities in the response surface of the number of pregnancies. Since the values of $\beta > 0$ are related to catastrophic events, estimating the catastrophe model and testing the statistical significance of the parameters $\alpha$ and $\beta$ is equivalent to empirically testing the presence of fertility catastrophic events.

The form of the probability density function of $\xi$ for stochastic catastrophe models (Equation \ref{eq:density} below) was developed by \citet{cobb1978stochastic, cobb1981parameter, cobb1983estimation, cobb1985applications, cobb2010estimation} as a multimodal generalization of distributions belonging to the exponential family, without relying on mixture densities \citep{cobb1983estimation}:
\begin{equation}
f(\xi; \alpha, \beta, \lambda, \sigma) = \frac{\phi}{\sigma^2} \exp
\left( 
\frac{\alpha (\xi - \lambda) + \frac{1}{2}  \beta (\xi - \lambda)^2 - \frac{1}{4}  (\xi - \lambda)^4  }{\sigma^2}
\right)
\label{eq:density}
\end{equation}
In Equation \ref{eq:density}, $\lambda$ and $\sigma$ are the location and scale parameters of the state variable $\xi$, respectively, and $\phi$ is a constant that normalizes the p.d.f. so that it has a unit integral over the range $\mathbb{R}$ \citep{cobb1981parameter}. Estimates of the parameters of interest $\alpha$ and $\beta$ can be obtained with information on the number of pregnancies on the basis of the maximum likelihood approach of \citet{cobb1985applications} augmented with the subspace fitting method of \citet{oliva1987gemcat}, in which $\xi$ is a canonical variable that results from a first-order polynomial approximation of measured dependent variables. Specifically, let $\mathbf{Y} = [\mathbf{1} \ \ | \ (Y_i)]$, $\mathbf{X} = [\mathbf{1} \ \ | \ (X_j)]$, 
$\mathbf{Z} = [\mathbf{1} \ \ | \  (Z_k)]$, and $\mathbf{w} = (\omega_0,\omega_1,\dots,\omega_p)'$, $\mathbf{a} =(a_0,a_1,\dots,a_q)'$, $\mathbf{b} = (b_0,b_1,\dots,b_r)'$. In this notation, $\mathbf{Y}$, $\mathbf{X}$, and $\mathbf{Z}$ are vectors that contain stochastic dynamics variables $Y_1, Y_2, \dots, Y_p$, $X_1, X_2, \dots, X_q$, and $Z_1, Z_2, \dots, Z_r$, respectively, and $\mathbf{1}$ is a vector of ones concatenated in the matrices $\mathbf{Y}$, $\mathbf{X}$, and $\mathbf{Z}$. Given this matrix notation:
\begin{equation}
\mathbf{\xi} = \mathbf{Yw}, \quad\mathbf{\alpha} = \mathbf{Xa}, \quad  \mathbf{\beta} = \mathbf{Zb},
\end{equation}
the optimization of the log-likelihood function,
\begin{equation}
\ell \left(\mathbf{a},\mathbf{b},\mathbf{w}; \mathbf{Y},\mathbf{X} \right) = \sum_{i \in D} \log \phi_i - \sum_{i \in D} \left(
\alpha_i \xi_i + \frac{1}{2} \beta_i \xi_i^2 - \frac{1}{4}\xi_i^4,
\right)
\label{eq:matloglike}
\end{equation}
with data $D$ ($i \in D$), provides estimates concerning the variables that explain the number of pregnancies during a fertility crisis and amidst fertility catastrophes. The likelihood function of equation \ref{eq:matloglike} does not necessarily have to integrate to one, as the likelihood is proportional to the probability distribution of the data given the parameters, but the normalization constants $\phi_i$ ensure that the probabilities integrate to one over the parameter space \citep{grasman2010fitting}.

If the null hypothesis $\mathbb{H}_0: \mathbf{b} = \mathbf{0}$ cannot be rejected at conventional significance levels, and the parameters $\mathbf{a}$ in $\mathbf{\alpha} = \mathbf{Xa}$ are statistically significant, then the fertility dynamics can be analyzed solely with traditional fertility models, without the nonlinear dynamics introduced by a catastrophic fertility shock. The statistical significance of the parameters $\mathbf{b}$ in $\mathbf{Zb}$ is a way to empirically test if catastrophes affected the number of pregnancies, since the parameters $\mathbf{b}$ are associated with catastrophic shocks $\mathbf{Z}$ that affect the dynamics of fertility through the control parameter $\mathbf{\beta} = \mathbf{Zb}$ in the catastrophe model.

In practice, catastrophic events are not limited to negative shocks, but can also capture positive effects---when the control parameters are equal to $\hat\beta > 0$, $\hat\alpha > 0$---, such as increases in fertility rates in a population explosion. Positive control parameters generate catastrophic distributions with support in the zero-to-infinite positive region ($\mathbb{R}^{[0,\infty^+)}$), adjacent to the right tail of the distribution of the number of pregnancies. Understanding the potential positive and negative impacts of catastrophic events can improve stakeholder coordination, guide stress testing exercises, and inform policymakers' decisions on emergency actions.

When the bifurcation parameter is positive ($\beta > 0$), catastrophic events $\mathbf{Z}$ affecting pregnancy rates lead to either (i) a sudden and sharp \textit{decrease} in the number of pregnancies if $\beta > 0$ and $\alpha < 0$, or (ii) an abrupt \textit{increase} in pregnancies when $\beta > 0$ and $\alpha > 0$. Through the parameters $\alpha$ and $\beta$, the fertility catastrophe model captures the multifactorial nature of fertility rate changes, with latent factors reflecting the varying importance of each factor across different regions and contexts. 

Among these factors, previous studies have linked fertility changes to increased female education, greater labor force participation by women, economic development, rising living standards, and urbanization. For instance, \citet{HumanFertility} highlighted the strong negative correlation between female education and fertility, suggesting that expanded career opportunities and delayed childbearing contribute to this relationship. \citet{GlobalFertility} discuss the societal shift toward smaller family preferences due to modernization and evolving social norms, as well as the rising cost of child-rearing—exacerbated by urbanization and higher living standards—leading to a contraction in family sizes. Furthermore, \citet{MiddleEastFertility} note that the trend of later marriages shortens the reproductive window, contributing to lower fertility rates.

Progress toward gender equality can also be seen as a contributing factor, as it grants women greater autonomy over their reproductive choices, often leading to delayed childbearing, which can impact fertility rates due to age-related biological factors \citep{GlobalFertility}. Both \citet{HumanFertility} and \citet{MiddleEastFertility} emphasize the increased availability and use of contraception as a key factor empowering women to control their fertility. Finally, \citet{MiddleEastFertility} highlight improvements in healthcare and reductions in child mortality rates as additional contributors to the global decline in fertility.

These factors interact in complex ways, with their relative importance varying across regions and contexts. This highlights the necessity of modeling specific drivers in different populations through the latent factors embedded in the control parameters $\alpha$ and $\beta$ of the fertility catastrophe model.

\section{A Cusp Catastrophe-Agent Based Mathematical Model of Teenage Pregnancy}\label{sec:ABM}

To introduce complexity into an agent-based model (ABM) of teenage pregnancy, we consider multi-level dynamics incorporating socioeconomic environments, peer-group pressure through social connections, and feedback mechanisms between these levels. Previous studies have proposed agent-based models of family planning \citep{o2023fpsim}, the low fertility trap \citep{kim2016agent}, and changes in fertility rates \citep{singh2016simulating}. Teenage pregnancy was modeled with ABM before by \citet{barroso2016unwanted}, but to the best of our knowledge, no ABM model was combined before with Thom's nonlinear equations of morphogenesis. 

\subsection{Agents and Social Networks}

Each agent \( A_i \) exists in a structured social network \( G = (V, E) \), where:
\begin{itemize}
    \item \( V \) is the set of agents (vertices).
    \item \( E \) is the set of edges representing social connections between agents.
\end{itemize}

Let \( \mathcal{N}_i \) denote the neighborhood of agent \( A_i \) in the network (i.e., the set of agents connected to \( A_i \)).

In addition to individual-level latent factors such as education, contraceptive use, and parental guidance, we introduce peer-group pressure through network interactions. Peer pressure affects behaviors such as contraceptive use or substance abuse.

Define \textbf{peer influence \( PI_i(t) \)} as a function of the behaviors of agent \( A_i \)'s neighbors in the network. For example, contraceptive use \( C_i(t) \) and substance abuse \( SAb_i(t) \) can be influenced by the agent’s neighborhood as follows:

\[
PI_i^{C}(t) = \frac{1}{|\mathcal{N}_i|} \sum_{j \in \mathcal{N}_i} C_j(t)
\]
\[
PI_i^{SAb}(t) = \frac{1}{|\mathcal{N}_i|} \sum_{j \in \mathcal{N}_i} SAb_j(t)
\]

Here, \( PI_i^{C}(t) \) is the average contraceptive use in agent \( A_i \)’s network, and \( PI_i^{SAb}(t) \) is the average level of substance abuse in the network.

Agents also exist within a broader \textbf{socio-economic environment}, modeled as a collection of interconnected communities or regions. Each environment has its own characteristics, which affect the agents within it.

Let \( \mathcal{R} \) denote the set of regions, and each agent \( A_i \) belongs to a region \( R_k \in \mathcal{R} \). Each region has the following characteristics:
\begin{itemize}
    \item \textbf{Regional Deprivation \( D_k(t) \)}: A continuous variable representing the degree of socioeconomic hardship in region \( R_k \) at time \( t \).
    \item \textbf{Regional Violence \( V_k(t) \)}: A binary variable representing the prevalence of violence in region \( R_k \) at time \( t \).
    \item \textbf{Regional Education Quality \( EQ_k \)}: A continuous variable representing the quality of education available in region \( R_k \).
\end{itemize}

These regional factors interact with the individual-level factors and peer-group dynamics.

The probability of pregnancy incorporates both social network effects and regional-level socio-economic environmental influences. Define the pregnancy probability as:

\[
P(S_i(t+1) = 1 \mid X_i(t)) = \frac{1}{
\begin{split}
1 + \exp \left[ -( \beta_0 + \beta_1 E_i + \beta_2 C_i(t) + \beta_3 P_i + \beta_4 PI_i^{C}(t) \right. \\ +  \gamma_1 D_i(t) + \gamma_2 V_i(t) + \gamma_3 SAb_i(t) + \gamma_4 PI_i^{SAb}(t) \\ \left. + \theta_1 D_k(t) + \theta_2 V_k(t) + \theta_3 EQ_k ) \right] \end{split} }
\]

Where:
\begin{itemize}
    \item \( \beta_0 \): Baseline risk of pregnancy.
    \item \( \beta_4, \gamma_4 \): Coefficients for peer-group influence on contraceptive use and substance abuse.
    \item \( \theta_1, \theta_2, \theta_3 \): Coefficients for regional deprivation, violence, and education quality, respectively.
\end{itemize}

Thus, the probability of pregnancy depends on:
\begin{itemize}
    \item \textbf{Latent factors}: Education \( E_i \), contraceptive use \( C_i(t) \), and parental guidance \( P_i \).
    \item \textbf{Peer influence}: \( PI_i^{C}(t) \) and \( PI_i^{SAb}(t) \), representing peer-group behaviors in contraceptive use and substance abuse.
    \item \textbf{Shocks and external factors}: Socioeconomic deprivation \( D_i(t) \), violence \( V_i(t) \), and substance abuse \( SAb_i(t) \).
    \item \textbf{Regional factors}: Deprivation \( D_k(t) \), violence \( V_k(t) \), and education quality \( EQ_k \).
\end{itemize}

To reflect feedback mechanisms in peer groups, agent behavior can evolve based on their neighbors’ behavior. The evolution of contraceptive use and substance abuse follows:

\[
C_i(t+1) = \sigma\left(\lambda_1 C_i(t) + \lambda_2 PI_i^{C}(t) + \lambda_3 P_i\right)
\]
\[
SAb_i(t+1) = \sigma\left(\nu_1 SAb_i(t) + \nu_2 PI_i^{SAb}(t) + \nu_3 D_i(t) + \nu_4 V_i(t)\right)
\]

Where \( \sigma(x) = \frac{1}{1 + \exp(-x)} \) is the sigmoid function, and \( \lambda_1, \lambda_2, \lambda_3, \nu_1, \nu_2, \nu_3, \nu_4 \) are parameters that define the impact of individual, peer, and shock factors on future contraceptive use and substance abuse.

Each region \( R_k \) evolves over time, with factors like deprivation and violence changing in response to systemic shocks, policy interventions, and feedback from the agent population. Regional deprivation and violence evolve as follows:

\[
D_k(t+1) = \rho_1 D_k(t) + \rho_2 \left(\frac{1}{|R_k|} \sum_{i \in R_k} S_i(t)\right) + \epsilon_D(t)
\]
\[
V_k(t+1) = \tau_1 V_k(t) + \tau_2 \left(\frac{1}{|R_k|} \sum_{i \in R_k} SAb_i(t)\right) + \epsilon_V(t)
\]

Where \( \epsilon_D(t) \) and \( \epsilon_V(t) \) are noise terms, and \( \rho_2 \) and \( \tau_2 \) reflect the feedback from the prevalence of pregnancy and substance abuse in the region to its overall socioeconomic deprivation and violence.

There are interactions between individual agents’ behavior and their regional environments. For example, an increase in regional violence may lead to an increase in peer-group pressure for risky behaviors, further deteriorating regional conditions. These feedback loops create \textit{emergent phenomena} at the system level.

To model these interactions:

\[
PI_i^{SAb}(t+1) = \sigma\left(\eta_1 \left(\frac{1}{|\mathcal{N}_i|} \sum_{j \in \mathcal{N}_i} SAb_j(t)\right) + \eta_2 V_k(t)\right)
\]

Where \( \eta_2 \) accounts for the influence of regional violence on peer-group behaviors, linking the regional-level shocks with individual behavior at the social network level.

The model tracks:
\begin{itemize}
    \item \textbf{Pregnancy rates}: Both at the individual level and across regions.
    \item \textbf{Behavioral evolution}: How contraceptive use and substance abuse evolve based on peer-group pressure and regional environments.
    \item \textbf{Regional dynamics}: Changes in socioeconomic deprivation, violence, and education quality over time.
\end{itemize}

This enhanced agent-based model creates a \textbf{multi-level feedback system} where teenage pregnancy, substance abuse, and other behaviors emerge from interactions between individuals, their peer networks, and the socio-economic environment. The complexity arises from the network effects, regional dynamics, and feedback loops within and between these levels.

Each agent \( A_i \) is part of a social network \( G \), with the following attributes:
\begin{itemize}
    \item \textbf{Pregnancy State} \( y_i(t) \): A continuous variable representing the pregnancy state of agent \( A_i \) at time \( t \).
    \item \textbf{Contraceptive Use} \( C_i(t) \): A binary variable indicating the contraceptive use by agent \( A_i \) at time \( t \).
    \item \textbf{Deprivation} \( D_i(t) \): A continuous variable representing the socioeconomic deprivation of agent \( A_i \).
    \item \textbf{Violence} \( V_i(t) \): A continuous variable representing the exposure to violence.
    \item \textbf{Substance Abuse} \( SAb_i(t) \): A binary variable representing substance abuse by agent \( A_i \) at time \( t \).
\end{itemize}

\subsection*{Cusp Catastrophe Control Parameters}

The dynamics of pregnancy evolution are governed by the cusp catastrophe potential function:
\[
V(y, \alpha, \beta) = \frac{1}{4} y^4 - \frac{1}{2} \alpha y^2 - \beta y
\]
where:
\begin{itemize}
    \item \( y \): State variable (pregnancy state) for the agent.
    \item \( \alpha \): Normal control parameter, affected by contraceptive use.
    \item \( \beta \): Catastrophic control parameter, affected by deprivation, violence, and substance abuse.
\end{itemize}

\subsection*{Normal Control Parameter \( \alpha_i(t) \)}

The normal control parameter \( \alpha_i(t) \) evolves based on contraceptive use:
\[
\alpha_i(t) = \alpha_0 + \lambda_1 C_i(t)
\]
where:
\begin{itemize}
    \item \( \alpha_0 \) is a baseline value for the normal control.
    \item \( \lambda_1 \) is the influence of contraceptive use on the normal control.
\end{itemize}

\subsection*{Catastrophic Control Parameter \( \beta_i(t) \)}

The catastrophic control parameter \( \beta_i(t) \) is influenced by socioeconomic deprivation, violence, and substance abuse:
\[
\beta_i(t) = w_1 D_i(t) + w_2 V_i(t) + w_3 SAb_i(t)
\]
where \( w_1, w_2, w_3 \) are weights reflecting the impact of deprivation, violence, and substance abuse, respectively.

The pregnancy state \( y_i(t) \) evolves according to the gradient of the potential function:
\[
\frac{dy_i(t)}{dt} = - \frac{\partial V}{\partial y} = - y_i^3 + \alpha_i(t) y_i(t) + \beta_i(t)
\]

Peer-group pressure influences contraceptive use and substance abuse. Let \( \mathcal{N}_i \) represent the set of neighbors of agent \( A_i \) in the social network \( G \). The influence on contraceptive use is modeled as:
\[
C_i(t+1) = \sigma\left( \lambda_2 \cdot \frac{1}{|\mathcal{N}_i|} \sum_{j \in \mathcal{N}_i} C_j(t) \right)
\]
where \( \sigma(x) \) is the sigmoid function.

Similarly, substance abuse evolves based on peer influence:
\[
SAb_i(t+1) = \sigma\left( \lambda_3 \cdot \frac{1}{|\mathcal{N}_i|} \sum_{j \in \mathcal{N}_i} SAb_j(t) \right)
\]

The overall simulation evolves by updating the pregnancy state \( y_i(t) \), contraceptive use \( C_i(t) \), and substance abuse \( SAb_i(t) \) iteratively. Each agent's state is influenced by the catastrophic control parameter \( \beta_i(t) \), as well as peer-group pressure from its neighbors.

\begin{figure}[ht]
\centering
 \includegraphics[width=14cm]{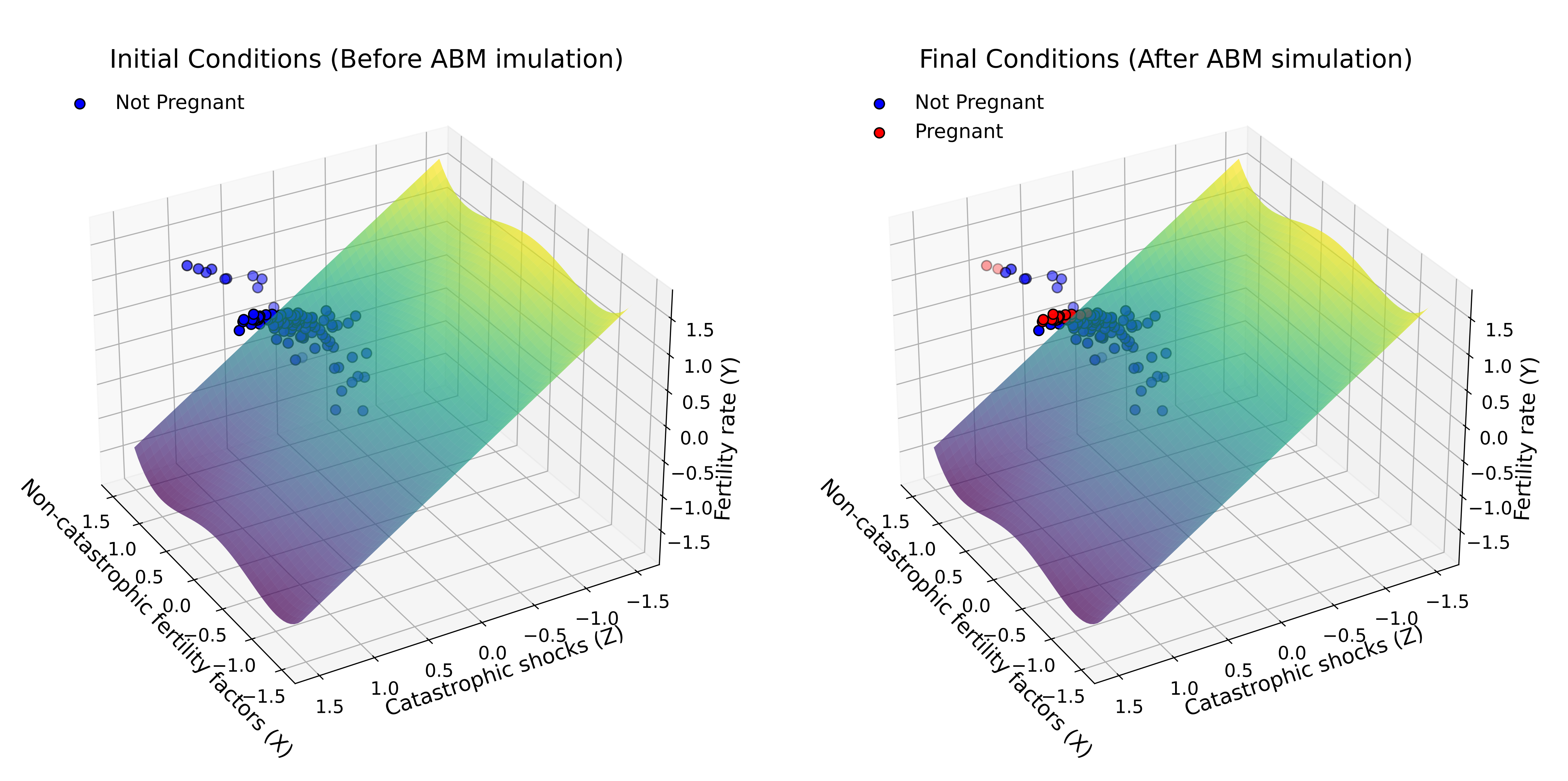} 
 \caption{Simulation of teenage pregnancy in an agent-based model using a cusp catastrophe surface. Left: all agents (represented as blue markers) are in a non-pregnant state, positioned according to non-catastrophic fertility factors (X-axis) and catastrophic shocks (Z-axis), such as deprivation and violence. Right: After the ABM simulation, some agents (red markers) have transitioned into a pregnant state, triggered by an increase in catastrophic shocks (Z-axis) or decreases in non-catastrophic factors (X-axis).}
\label{fig2}
\end{figure}

Figure \ref{fig2} shows simulations of the Agent-Based Model (ABM) applied to teenage pregnancy using the cusp catastrophe model. The model incorporates both non-catastrophic fertility factors---such as education and contraceptive use---and ---catastrophic shocks---deprivation, violence, and substance abuse. Figure \ref{fig2} illustrate the state of agents (teenagers) before and after the simulation, showing who became pregnant under varying socio-economic conditions.

In the initial conditions (Figure \ref{fig2} left), all agents (represented by blue markers) are in the non-pregnant state at the beginning of the simulation. The non-catastrophic fertility factors (X-axis) includes variables like education quality, parental guidance, and contraceptive use, which are generally preventive against teenage pregnancy. The catastrophic shocks (Z-axis), in turn, reflects negative external factors such as socioeconomic deprivation, violence, and substance abuse, which are not immediately strong enough to trigger pregnancy in the initial state. The Y-axis represents the fertility rate, which is initially low or neutral for all agents, indicating stability in terms of pregnancy prevention.

Figure \ref{fig2} (left) highlights a relatively stable system where the agents are unlikely to get pregnant due to a combination of effective contraceptive use and low exposure to catastrophic factors. According to the catastrophe theory, the agents are positioned on the stable part of the cusp surface, where minor changes in conditions would not result in significant shifts in pregnancy status.

However, after the ABM Simulation (Figure \ref{fig2} right), some teenagers (agents represented by red markers) transition into the pregnant state. These agents experienced a catastrophic shift due to the cumulative effects of deprivation, violence, and substance abuse, pushing them over the cusp. The cusp catastrophe model shows a bifurcation region where small changes in the control parameters (non-catastrophic factors and catastrophic shocks) can lead to abrupt and significant changes in the pregnancy state (Y-axis). This explains why some agents, initially stable, suddenly became pregnant, caused, for example, by peer influence through social networks. Some agents may have shifted into the pregnant state due to feedback from their peers, where behaviors such as substance abuse or lack of contraceptive use spread through social groups, amplifying the impact of catastrophic factors. Violence, deprivation, and substance abuse play a roles as catastrophic shocks (Z-axis), increasing over time in the simulation the risk of pregnancy for some agents, leading to a bifurcation where these individuals cross a critical threshold and become pregnant. The steep decline in the cusp surface reflects this catastrophic transition.

Figure \ref{fig2} (right) shows the system transitioning into a more complex and dynamic state, with some agents falling into pregnancy due to the interaction of catastrophic socio-economic shocks and latent factors. The cusp catastrophe model explains how these agents shift from the stable region (non-pregnant) to the unstable region (pregnant) when critical control parameters (such as deprivation or substance abuse) surpass threshold values.

The cusp catastrophe dynamics effectively illustrates the non-linear nature of fertility dynamics. In reality, teenagers may remain in a non-pregnant state despite small fluctuations in environmental conditions, but once critical factors (like poverty or violence) reach a tipping point, pregnancy occurs suddenly and unpredictably. This illustrates the catastrophic nature of teenage pregnancy in vulnerable socioeconomic environments, where a combination of negative factors can lead to dramatic and often irreversible consequences.

In terms of policy implications, the results highlight the need for targeted interventions aimed at mitigating catastrophic shocks like socioeconomic deprivation and violence, which, according to the model, play a critical role in pushing teenagers toward pregnancy. Preventive measures such as improving access to contraceptives and educational programs, can help maintain the system in a stable, low-pregnancy state.

\section{Discussion}\label{sec:discussion}

A fertility catastrophe model---based on Thom's nonlinear differential equations of morphogenesis---offers a novel framework for understanding reproductive health dynamics across macrodemographic, mesodemographic, and microdemographic levels. This model provides a structured approach to examining both gradual and abrupt shifts in fertility patterns, emphasizing how systemic latent factors interact with sudden perturbations to produce non-linear demographic transitions.

At the macrodemographic level, the model conceptualizes a three-dimensional catastrophe surface, mapping the interplay between latent fertility determinants—such as economic structures, cultural norms, and policy frameworks—and catastrophic shocks, including financial crises, pandemics, and armed conflicts. This analytical perspective is particularly relevant in understanding how small, incremental changes in socioeconomic conditions can, under specific thresholds, culminate in abrupt fertility declines. Japan’s persistent fertility decline exemplifies such a dynamic, wherein long-term socioeconomic transformations—rising costs of child-rearing, shifts in gender norms, and labor market rigidities—intersect with exogenous shocks to reinforce sub-replacement fertility levels. Similarly, the COVID-19 pandemic that induced a global baby bust \citep{kearney2023us} contributed to a significant contraction in birth rates across numerous advanced economies due to its related disruptions in economic stability, healthcare accessibility, and work-life balance. The integration of catastrophe theory with fertility analysis suggests that while structural factors drive long-term demographic trends, tipping points triggered by acute crises can lead to sudden demographic shifts that are difficult to reverse.

At the mesodemographic level, the model serves as a powerful tool for dissecting fertility patterns within specific communities, bridging the gap between national demographic trends and individual reproductive behaviors. Fertility is not uniform across geographic and social contexts; rather, it is shaped by local socioeconomic conditions, cultural expectations, and access to reproductive healthcare. Regional disparities in fertility rates, evident within countries with high spatial heterogeneity, highlight the relevance of meso-level analysis. For instance, intra-national fertility variations in the United States or sub-Saharan Africa can be attributed to differing levels of economic development, healthcare infrastructure, and social capital. By examining the interaction between local latent factors and potential catastrophic shocks, the model facilitates targeted policy interventions that mitigate fertility crises in vulnerable communities. Furthermore, the model is instrumental in analyzing the fertility consequences of migration. Shifts in population composition due to migration alter social dynamics, influencing reproductive behaviors within host communities. In this context, migration serves as both a driver of fertility adaptation and a mechanism for demographic stabilization in regions experiencing population decline.

At the microdemographic level, the model provides insight into fertility decision-making at the individual and household levels, particularly in contexts of heightened vulnerability. A compelling application of catastrophe theory is the cusp catastrophe-agent-based model of teenage pregnancy, which illustrates how latent factors—such as access to sex education, parental guidance, and peer influence—interact with catastrophic shocks like socioeconomic deprivation and exposure to violence. This model elucidates how minor perturbations in an adolescent’s environment can precipitate disproportionately large shifts in reproductive outcomes, underscoring the non-linear nature of fertility transitions. The findings align with empirical research demonstrating that individual-level exposure to instability, such as armed conflict, can alter fertility preferences and behaviors. For instance, high-resolution georeferenced data from 25 sub-Saharan African countries, combined with records of violent events from the Armed Conflict Location and Event Data Project (ACLED) and fertility data from the Demographic and Health Surveys, reveal that localized exposure to armed conflict is associated with reductions in both preferred family size and recent childbearing \citep{thiede2020exposure}. However, while such effects are pronounced at the individual level, they do not always translate into macrodemographic shifts when averaged across entire populations. This discrepancy highlights the necessity of multi-scalar analysis, recognizing that catastrophic effects may manifest distinctly at different demographic levels.

Existing fertility research in advanced societies, as reviewed by \citet{balbo2013fertility}, reinforces the importance of multi-level determinants. At the macro level, economic cycles, policy measures, technological advancements, and cultural shifts influence fertility rates, with recessions and economic uncertainty often delaying childbearing, whereas policies such as parental leave and subsidized childcare support fertility recovery. At the meso level, fertility is shaped by social interactions, urbanization patterns, and social capital, as peer influences, kin networks, and residential contexts play a fundamental role in reproductive decisions. At the micro level, individual agency, partnership stability, gender roles, and economic security directly shape fertility outcomes. While traditional demographic models emphasize the cumulative impact of these determinants, catastrophe theory introduces an alternative lens, suggesting that fertility systems can undergo sudden, irreversible transitions when critical thresholds are exceeded. This perspective is particularly salient in explaining the demographic shocks induced by events such as wars, economic depressions, and pandemics, as evidenced by the COVID-19-induced fertility downturn \citep{kearney2023us}.

The implications of this framework extend beyond theoretical demography, offering practical insights for policymakers and researchers. Understanding the conditions under which gradual fertility changes escalate into catastrophic declines is crucial for designing resilient reproductive health policies. Empirical studies integrating catastrophe modeling with demographic data can illuminate early-warning signals for fertility crises, enabling proactive interventions. Furthermore, interdisciplinary collaborations—drawing from systems biology, complexity theory, and behavioral economics—can refine the model’s predictive capabilities, fostering a more nuanced understanding of fertility dynamics in an era of heightened global uncertainty.

\section{Conclusion}\label{sec:conc}

The integration of Thom's nonlinear morphogenesis equations with fertility theory offers a groundbreaking perspective on reproductive health, providing a more comprehensive understanding of both stable fertility patterns and abrupt fertility transitions. Insights from this model can inform more effective public health policies that account for both long-term socioeconomic trends and potential catastrophic impacts on fertility. Additionally, the model can help evaluate the effectiveness of local policies and programs aimed at influencing fertility rates, such as family planning initiatives and economic development efforts. By analyzing how these policies interact with existing latent factors and potential shocks, policymakers can gain a deeper understanding of their impact on reproductive health within specific communities. The model's bifurcation set analysis also offers a visual representation of potential pathways and critical thresholds in fertility dynamics. This analysis illustrates how minor shifts in underlying socioeconomic and environmental conditions can lead to significant changes in fertility rates. Identifying these critical points provides policymakers with valuable insights into potential turning points, allowing them to implement targeted interventions in reproductive health.

By synthesizing catastrophe theory with contemporary demographic research, we proposed a more integrative and dynamic understanding of fertility transitions was proposed. Our approach underscores the necessity of multi-level approaches in demographic analysis and emphasizes the critical role of catastrophic exogenous shocks in shaping short-term and long-term fertility trajectories. Future research can refine the empirical applications of catastrophe modeling in fertility studies, exploring its potential to forecast demographic tipping points and inform evidence-based policy responses.

\bibliography{IPC2025}

\appendix

\section{Python Code of to replicate the simulations of the Cusp Catastrophe-Agent Based Mathematical Model of Teenage Pregnancy}

\begin{scriptsize}
\begin{lstlisting}[language=Python]

%\begin{minted}[frame=lines, bgcolor=lightgray, linenos]{python}
import numpy as np
import matplotlib.pyplot as plt
from mpl_toolkits.mplot3d import Axes3D

# Parameters for Cusp Catastrophe model
alpha_0 = 1.5  # Reduced base alpha to make pregnancy slightly more likely
lambda_1 = 0.8  # Impact of contraceptive use
w1, w2, w3 = 0.8, 0.6, 0.4  # Weights for deprivation, violence, and substance abuse

# Time steps and number of agents
n_steps = 100
n_agents = 100

# Create neighborhoods with different socio-economic conditions
n_neighborhoods = 5
# Each neighborhood has [deprivation, violence, education_quality]
neighborhoods = np.random.uniform(0, 1, size=(n_neighborhoods, 3))  

# Initialize agents with attributes: pregnancy state, contraceptive use, 
# deprivation, violence, substance abuse
agents = {
    # Initial pregnancy state (y)
    'pregnancy_state': np.random.normal(0, 0.1, size=n_agents),  
    # More use contraception
    'contraceptive_use': np.random.choice([0, 1], size=n_agents, p=[0.7, 0.3]),  
    'deprivation': np.random.uniform(0, 1, size=n_agents),  # Deprivation level
    'violence': np.random.uniform(0, 1, size=n_agents),  # Violence level
    # Fewer engage in substance abuse
    'substance_abuse': np.random.choice([0, 1], size=n_agents, p=[0.85, 0.15])  
}

# Sigmoid function for logistic update of contraceptive use and substance abuse
def sigmoid(x):
    return 1 / (1 + np.exp(-x))

# Function to compute the catastrophic control parameter beta
def compute_beta(deprivation, violence, substance_abuse):
    return w1 * deprivation + w2 * violence + w3 * substance_abuse

# Simulation function
def simulate_agents(agents, n_steps):
    fertility_status = np.zeros(n_agents)  # Track final pregnancy status
    
    for step in range(n_steps):
        for i in range(n_agents):
            # Compute normal control parameter alpha
            alpha_i = alpha_0 + lambda_1 * agents['contraceptive_use'][i]
            
            # Compute catastrophic control parameter beta
            beta_i = compute_beta(agents['deprivation'][i], 
                                  agents['violence'][i], 
                                  agents['substance_abuse'][i])
            
            # Update pregnancy state (cusp catastrophe dynamics)
            dy_dt = -agents['pregnancy_state'][i]**3 + alpha_i *agents['pregnancy_state'][i] 
            + beta_i
            agents['pregnancy_state'][i] += dy_dt * 0.01  # Time step update
            
            # Update contraceptive use and substance abuse with some randomness
            # More likely to use contraception
            agents['contraceptive_use'][i] = sigmoid(np.random.uniform(0, 1)) > 0.7
            # Less likely to abuse substances
            agents['substance_abuse'][i] = sigmoid(np.random.uniform(0, 1)) > 0.85 
        
        # Determine fertility status based on pregnancy state
        # Adjust threshold for pregnancy
        fertility_status = np.array([1 if sigmoid(y) > 0.8 else 0 for y in 
        agents['pregnancy_state']])  
    
    return fertility_status

# Run the simulation
initial_fertility_status = np.zeros(n_agents)  # Assume no one is pregnant initially
final_fertility_status = simulate_agents(agents, n_steps)

# Cusp catastrophe surface
x = np.linspace(-1.5, 1.5, 50)  # Non-catastrophic fertility factors
z = np.linspace(-1.5, 1.5, 50)  # Catastrophic shocks
x, z = np.meshgrid(x, z)
y = 0.25 * x**4 - 0.5 * x**2 - z

# Plotting the surface and results
fig = plt.figure(figsize=(18, 8))

# Initial state (Before simulation)
ax1 = fig.add_subplot(121, projection='3d')

# Plot the cusp catastrophe surface
ax1.plot_surface(x, z, y, cmap='viridis', alpha=0.7, edgecolor='none')

# Overlay the agents' initial states (no pregnancies initially)
ax1.scatter(agents['contraceptive_use'], 
           compute_beta(agents['deprivation'], 
                        agents['violence'], 
                        agents['substance_abuse']),
           agents['pregnancy_state'], 
           c='blue', label='Not Pregnant', s=50, edgecolor='k')

ax1.view_init(elev=30, azim=150)  # Set the view angle
ax1.set_title('Initial Conditions (Before ABM imulation)', fontsize=18)
ax1.set_xlabel('Non-catastrophic fertility factors (X)', fontsize=15)
ax1.set_ylabel('Catastrophic shocks (Z)', fontsize=15)
ax1.set_zlabel('Fertility rate (Y)', fontsize=15)
ax1.legend(loc='upper left', fontsize=14, frameon=False) 

# Increase the size of ticks
ax1.tick_params(axis='both', which='major', labelsize=13)

# Final state (After simulation)
ax2 = fig.add_subplot(122, projection='3d')

# Plot the cusp catastrophe surface
ax2.plot_surface(x, z, y, cmap='viridis', alpha=0.7, edgecolor='none')

# Separate the agents by pregnancy status
pregnant_agents = final_fertility_status == 1
non_pregnant_agents = final_fertility_status == 0

# Overlay the agents' final states (color-coded by pregnancy status)
ax2.scatter(agents['contraceptive_use'][non_pregnant_agents], 
           compute_beta(agents['deprivation'][non_pregnant_agents], 
                        agents['violence'][non_pregnant_agents], 
                        agents['substance_abuse'][non_pregnant_agents]),
           agents['pregnancy_state'][non_pregnant_agents], 
           c='blue', label='Not Pregnant', s=50, edgecolor='k')

ax2.scatter(agents['contraceptive_use'][pregnant_agents], 
           compute_beta(agents['deprivation'][pregnant_agents], 
                        agents['violence'][pregnant_agents], 
                        agents['substance_abuse'][pregnant_agents]),
           agents['pregnancy_state'][pregnant_agents], 
           c='red', label='Pregnant', s=50, edgecolor='k')

ax2.view_init(elev=30, azim=150)  # Set the view angle
ax2.set_title('Final Conditions (After ABM simulation)', fontsize=18)
ax2.set_xlabel('Non-catastrophic fertility factors (X)', fontsize=15)
ax2.set_ylabel('Catastrophic shocks (Z)', fontsize=15)
ax2.set_zlabel('Fertility rate (Y)', fontsize=15)
ax2.legend(loc='upper left', fontsize=14, frameon=False)  

# Increase the size of ticks
ax2.tick_params(axis='both', which='major', labelsize=13)

# Adjust layout to reduce the space between figures
plt.subplots_adjust(left=0.05, right=0.95, top=0.9, bottom=0.1, wspace=-0.05)

# Save the figure at 660 DPI
plt.savefig("cusp_catastrophe_simulation.png", dpi=660)
plt.show()
%\end{minted}    
\end{lstlisting}
\end{scriptsize}

\end{document}